\documentclass[conference]{IEEEtran}
\IEEEoverridecommandlockouts
\usepackage{cite}
\usepackage{amssymb,amsfonts}
\usepackage[tbtags]{amsmath}
\usepackage{algorithmic}
\usepackage{graphicx}
\usepackage{subcaption}
\usepackage{textcomp}
\usepackage{xcolor}
\usepackage{soul}
\usepackage{url}
\usepackage{stfloats}
\usepackage{makecell}
\usepackage{mathtools}
\usepackage{float}
\usepackage{gensymb}
\usepackage{booktabs}
\usepackage[export]{adjustbox}
\def\BibTeX{{\rm B\kern-.05em{\sc i\kern-.025em b}\kern-.08em
    T\kern-.1667em\lower.7ex\hbox{E}\kern-.125emX}}

\newcommand{\bpsi}{\boldsymbol{\Psi}}

\begin{document}


\title{Trade-Off Between Beamforming and Macro-Diversity Gains in Distributed mMIMO}

\author{Eduardo N. Tominaga\IEEEauthorrefmark{1}, Hsuang-Jung Su\IEEEauthorrefmark{2}\IEEEauthorrefmark{3}, Jinfeng Du\IEEEauthorrefmark{3}, Sivarama Venkatesan\IEEEauthorrefmark{3}, Richard D. Souza\IEEEauthorrefmark{4}, Hirley Alves\IEEEauthorrefmark{1}\\

\IEEEauthorblockA{
		\IEEEauthorrefmark{1}6G Flagship, Centre for Wireless Communications (CWC), University of Oulu, Finland.\\
		Email: \{eduardo.noborotominaga, hirley.alves\}@oulu.fi\\
        \IEEEauthorrefmark{2}Graduate Institute of Communication Engineering, National Taiwan University, Taipei, Taiwan. Email: hjs@ntu.edu.tw \\
        \IEEEauthorrefmark{3}Nokia Bell Labs, Murray Hill, NJ, 07974, USA. Email: \{jinfeng.du,venkat.venkatesan\}@nokia-bell-labs.com.\\
		\IEEEauthorrefmark{4}Federal University of Santa Catarina (UFSC), Florian\'{o}polis, Brazil. Email: richard.demo@ufsc.br\\
	}
}

\maketitle

\begin{abstract}
Industry and academia have been working towards the evolution from Centralized massive Multiple-Input Multiple-Output (CmMIMO) to Distributed mMIMO (DmMIMO) architectures. Instead of splitting a coverage area into many cells, each served by a single Base Station equipped with several antennas, the whole coverage area is jointly covered by several Access Points (AP) equipped with few or single antennas. Nevertheless, when choosing between deploying more APs with few or single antennas or fewer APs equipped with many antennas, one observes an inherent trade-off between the beamforming and macro-diversity gains that has not been investigated in the literature. Given a total number of antenna elements and total downlink power, under a channel model that takes into account a probability of Line-of-Sight (LoS) as a function of the distance between the User Equipments (UEs) and APs, our numerical results show that there exists a ``sweet spot" on the optimal number of APs and of antenna elements per AP which is a function of the physical dimensions of the coverage area.
\end{abstract}

\begin{IEEEkeywords}
Distributed massive MIMO, beamforming gain, macro-diversity gain, and spectral efficiency.
\end{IEEEkeywords}

\section{Introduction}

\par Massive Multiple-Input Multiple-Output (mMIMO) is one of the major physical layer technologies introduced in the Fifth Generation (5G) of wireless communications networks \cite{agiwal2016}. It features Base Stations (BS) equipped with many antenna elements, thus providing them with very high beamforming gains and spatial multiplexing capabilities \cite{emil2017}.

\par However, the current 5G deployments still adopt the cellular network paradigm; thus, the traditional Centralized mMIMO (CmMIMO) approach does not solve the issues of inter-cell interference and unequal performance between User Equipments (UEs) located at the cell center and UEs located at the cell edges \cite{matthaiaou2021}. Aiming to solve the aforementioned issues, the research community has been working towards evolving from CmMIMO to Distributed mMIMO (DmMIMO), also known as Cell-Free mMIMO. Instead of having a single BS in each cell equipped with several antennas, the coverage area is served by multiple Access Points (APs), each equipped with few or single antenna elements, and connected to a common Central Processing Unit (CPU) through fronthaul connections. This approach may provide uniform wireless coverage and solve the inter-cell interference problem via smart AP clustering and interference mitigation schemes \cite{demir2021}.

\par Most of the works on DmMIMO advocate for its significant performance improvements compared to CmMIMO in terms of data rates and uniform wireless coverage, e.g., \cite{interdonato2019,emil2019,fang2021}. By deploying several APs, one obtains macro-diversity gains that guarantee a more uniform wireless coverage. However, since the APs are equipped with few or a single antenna, they do not present the same beamforming gains or spatial multiplexing capabilities of a single BS equipped with several antennas. Thus, given the total number of antenna elements, a trade-off between macro-diversity and beamforming gains has been studied in some works. In the case of indoor industrial scenarios, this trade-off was studied in \cite{choi2020,alonzo2020}. Based on the results of a measurement campaign, the authors in \cite{choi2020} found that semi-distributed setups (i.e., setups that adopt few APs equipped with multiple antenna elements) present a performance comparable to that of the fully distributed setups (i.e., setups that adopt many single antenna APs) in terms of achievable downlink Spectral Efficiency (SE), with the advantage of presenting lower deployment cost. However, their setup had only eight antenna elements. In \cite{alonzo2020}, authors compared centralized, partially distributed, and fully distributed mMIMO setups regarding block error rates. They found that the DmMIMO setups only provide strong performance gains when the beamformers can properly handle inter-user interference. For micro-urban outdoor scenarios, the trade-off between the number of APs and antenna elements per AP was studied in \cite{ngo2018,ito2022,femenias2020}. In \cite{ngo2018}, the authors investigated the  energy efficiency of DmMIMO systems. They fixed the total number of antenna elements and the network's total transmit power for a fair comparison. They found that the optimal number of antennas per AP depends on system parameters such as the coverage area's target spectral efficiency and size. Nevertheless, their work did not thoroughly investigate the relation between such parameters and the optimal number of APs. The authors in \cite{ito2022} observe that semi-distributed deployments have almost the same performance as the fully distributed deployments in uplink and downlink, with the advantage of requiring a much lower number of APs. Nonetheless, the optimal number of APs or antennas per AP as a function of any system parameter is not discussed. In \cite{femenias2020}, the authors showed that increasing the number of APs in a square coverage area with a side length of $1$ km, given a total number of antenna elements and number of users, leads to higher mean per-user rates.

\par Inspired by the aforementioned works, especially \cite{ngo2018}, in this paper, we evaluate the downlink performance of a DmMIMO network. Fixing the total number of antenna elements and the total downlink transmit power budget, we investigate the ``sweet spot" in terms of the number of APs and antenna elements per AP. In other words, we investigate the trade-off between macro-diversity gain and beamforming gain. We conduct our analysis considering the impact of imperfect Channel State Information (CSI). Resorting to Monte Carlo simulations, our results show that, given a total number of antenna elements, the optimal numbers of APs and antenna elements per AP is a function of the dimensions of the coverage area. In other words, we found that there is an optimal density of APs (i.e., number of APs per $\text{km}^2$) that maximizes the mean per-user achievable SE. Such finding has not been reported in related works. We show that, for small coverage areas, having few APs equipped with many antennas yields better performance than having many APs equipped with few antennas. Nevertheless, as the coverage area increases and having the users uniformly distributed, having more APs in the system becomes increasingly advantageous.

\par This paper is organized as follows. The considered system model is introduced in Section \ref{systemModel}. 
The channel estimation methods studied in this work are described in Section \ref{channelEstimation}. Section \ref{downlinkDataTransmission} presents the downlink data transmission and the adopted performance metric. Section \ref{numericalResults} presents the numerical results based on Monte Carlo simulations. Finally, we conclude this work in Section \ref{Conclusions}.

\section{System Model}
\label{systemModel}

\par We consider a square coverage area with dimensions $l\times l\text{ m}^2$, where $Q$ APs, indexed by $q\in\{1,\ldots,Q\}$, cooperate to serve $K$ single-antenna UEs, which are indexed by $k\in\{1,\ldots,K\}$. Each AP has a Uniform Linear Array (ULA) containing $S$ antenna elements. The antenna spacing is half-wavelength, $d_H=1/2$.
The total number of antenna elements exceeds the number of UEs, $M=QS$, $M>K$.

\par
\begin{figure}[t]
    \centering
    \includegraphics[scale=0.5]{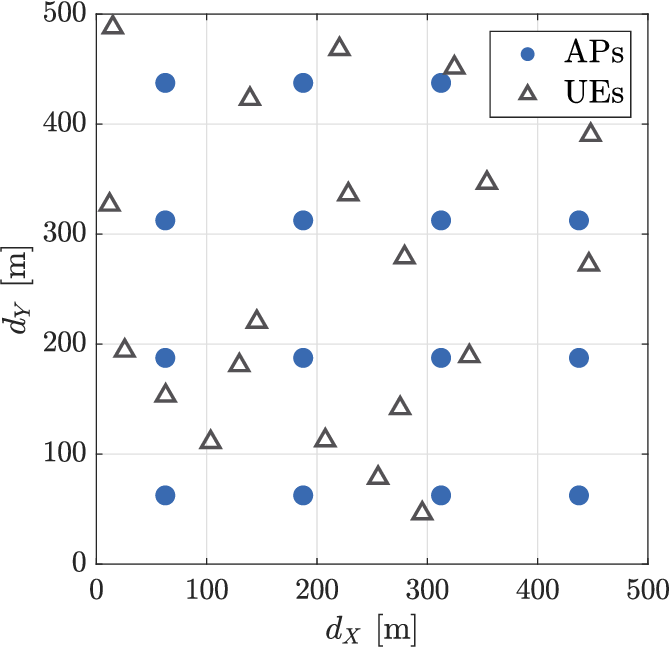}
    \caption{Illustration of the considered DmMIMO network for $Q=16$, $K=20$ and $l=500$ m.}
    \label{figSystemModel}
\end{figure}

 The $Q$ APs are distributed on a uniform grid of $\sqrt{Q}\times\sqrt{Q}$ APs, thus the value of $Q$ is selected such that $\sqrt{Q}$ is an integer. On the other hand, the $K$ UEs are uniformly distributed on the square coverage area. Denoting $(x_k,y_k)$ the coordinates of the location of the $k$-th UE, we have $x_k,y_k\sim\mathcal{U}(0,l)$. The considered system model is illustrated in Fig. \ref{figSystemModel} for $Q=16$ APs and $K=20$ UEs.

\par We assume a fully centralized operation, where each AP acts only as a Remote-Radio Head (RRH) and is connected to a common CPU through a fronthaul link. The APs forward the received signal samples to the CPU in the uplink and coherently transmit the signals generated by the CPU in the downlink. Moreover, all APs jointly serve all UEs.

\par We consider a system with bandwidth $B$ that operates at the carrier frequency $f_c$. By employing a multicarrier modulation scheme, the frequency resources are split into multiple flat-fading subcarriers. We assume the channel coefficients to be constant and frequency flat in a coherence block of length $\tau_c$ channel uses, and the independent block fading model. The performance analysis is carried out by studying a single statistically representative subcarrier during a coherence block~\cite{emil2017}. 


\par We assume channel reciprocity, i.e., the channel coefficients are the same during uplink and downlink transmissions. This assumption allows Time-Division Duplexing (TDD) operation. The CSI is acquired in the uplink by using pilot sequences, and it is utilized for coherent uplink receive combining and downlink transmit precoding \cite{emil2017}. The uplink training phase occupies $\tau_p$ samples, followed by a time instant reserved for channel estimation and processing tasks. The downlink data transmission phase occupies the next $\tau_d$ time instants. Thus, we have $\tau_c=\tau_p+1+\tau_d$. Note that we can have up to $\tau_p$ orthogonal pilot sequences.


\par We adopt the spatially correlated Rician fading channel model from 3GPP for MIMO simulations \cite{tr25.996} that was also used in \cite{ozdogan2019_2}. This model accounts for a Line-of-Sight (LoS) probability which is a function of the distance between the UE and the AP. 
The collective vector of wireless channel coefficients between the $k$-th UE and the $Q$ APs is
\begin{equation}
    \textbf{h}_k=[\textbf{h}_{k1}^T,\textbf{h}_{k2}^T,\ldots,\textbf{h}_{kQ}^T]^T\in\mathbb{C}^{M\times1},
\end{equation}
where $\textbf{h}_{kq}\in\mathbb{C}^{S\times1}$ is the vector of wireless channel coefficients between the $k$-th UE and the $q$-th AP such that
\begin{equation}
    \textbf{h}_{kq}\sim\mathcal{CN}(\Bar{\textbf{h}}_{kq},\textbf{R}_{kq}),
\end{equation}
where $\Bar{\textbf{h}}_{kq}\in\mathbb{C}^{S\times 1}$ is the LoS component and $\textbf{R}_{kq}\in\mathbb{C}^{S\times S}$ is the positive semidefinite covariance matrix describing the spatial correlation of the Non-Line-of-Sight (NLoS) components. The covariance matrices $\textbf{R}_{kq},\;\forall k,\;\forall q$ are assumed to be perfectly known.


\par We assume all UEs have the same fixed uplink transmit power $p$.  For downlink transmissions, the system has a maximum total transmit power denoted by $P$, which is assumed to be split equally among all the APs so that the per AP transmit power is
    $p_d=P/Q$.

\section{Channel Estimation}
\label{channelEstimation}

\par To obtain estimates of the wireless channel coefficients, the system adopts a set of mutually orthogonal pilot sequences $\psi_1,\psi_2,\ldots,\psi_{\tau_p}$ with length $\tau_p$, and with $\lVert\psi_t\rVert^2=\tau_p, t\in\{1,\ldots,\tau_p\}$. Herein we assume that $\psi_t$ is a column of $\sqrt{\tau_p}\textbf{I}_{\tau_p}\forall t$, i.e.,
    $\psi_t=\sqrt{\tau_p} \,\,[\textbf{I}_{\tau_p}]_t,\forall t\in\{1,\ldots,\tau_p\}$.


\par During a pilot transmission phase, the UEs transmit the pilot sequences with power $p$. We assume a crowded network such that $K>\tau_p$, thus, multiple UEs are assigned with the same pilot sequence, i.e., there is pilot contamination in the system.

\par We adopt a balanced random pilot assignment scheme. Even though this scheme is suboptimal, its performance is still substantially better than that of a purely random pilot assignment scheme \cite{femenias2019}. Each UE is allocated a pilot sequence, i.e., sequentially and cyclically selected from the set of available orthogonal pilot sequences. The index of the time instant allocated for the transmission of the pilot signal of UE $k$ is $n_k\in\{1,\ldots,\tau_p\}$. This index also corresponds to the index of the pilot sequence assigned to the $k$-th UE and is 
    $n_k=k-\left\lfloor\frac{k-1}{\tau_p}\right\rfloor \tau_p$.
The subset of UEs that use the same pilot sequence as the $k$-th UE is defined as $\mathcal{P}_k=\{i:n_i=n_k\}\subset\{1,\ldots,K\}$. 

\par During the pilot transmission phase, the received signal vector for the $q$-th AP at the time instant $n_k\in\{1,\ldots,\tau_p\}$ is $\textbf{y}_q^{\text{pilot}}[n_k]\in\mathbb{C}^{S\times 1}$,
\begin{equation}
        \textbf{y}_q^{\text{pilot}}[n_k]=\sum_{i\in\mathcal{P}_k}\sqrt{p_i}\,\, \textbf{h}_{iq} + \textbf{z}_q[n_k].
\end{equation}

\par The channel estimation takes place at time instant $\lambda=\tau_p+1$. We assume that the CPU has perfect statistical information of the channels, i.e., the correlation matrices $\textbf{R}_{kq}\;\forall k,\forall q$ are assumed to be perfectly known \cite{emil2017}.

\par Using the Linear Minimum Mean Square Error (LMMSE) estimation, the estimate $\hat{\textbf{h}}_{kq}[\lambda]$ of the channel coefficient $\textbf{h}_{kq}[n_k]$ can be computed by each AP according to \cite{zheng2022_journal}
\begin{align}
    \label{channelEstimates}
    \hat{\textbf{h}}_{kq}[\lambda] &= \sqrt{p_k} \,\, \textbf{R}_{kq}\,\bpsi_{n_kq}^{-1} \,\, \textbf{y}_{n_kq}^{\text{pilot}},  \\
    \label{eq:corr-matrix}
    \bpsi_{n_kq} &= \sum_{i\in\mathcal{P}_k} \,\, p_i \,\, \textbf{R}_{iq} + \sigma^2\textbf{I}_S
\end{align}
where \eqref{eq:corr-matrix} is the correlation matrix of the received signal.
%
%
\section{Downlink Data Transmission}
\label{downlinkDataTransmission}

\par In this work, we consider coherent downlink transmission, i.e., all the APs simultaneously transmit the same data symbol to a given UE. The received downlink signal at the $k$-th UE and at time instant $n\in\{\lambda+1,\ldots,\tau_c\}$ is
\begin{equation}
    y_k^{\text{dl}}[n]=\sum_{q=1}^Q \textbf{h}_{kq}[n]^H\textbf{x}_q[n] + z_k[n],
\end{equation}
where $z_k[n]\sim\mathcal{CN}(0,\sigma^2)$ is the AWGN sample at the receiver of the $k$-th UE,
\begin{equation}
    \textbf{x}_q[n]=\sqrt{p_d\mu_q}\sum_{k=1}^K \textbf{w}_{kq}[\lambda]s_k[n]
\end{equation}
is the signal transmitted by the $q$-th AP, $p_d$ is the maximum transmit power of each AP, $\mu_q$ is the normalization parameter for the precoding, $\textbf{w}_{kq}[\lambda]$ is the precoding vector utilized by the $q$-th AP for the transmission of the signal intended for the $k$-th UE, and $s_k[n]\sim\mathcal{CN}(0,1)$ is the data symbol intended for the $k$-th UE. Since the beamforming vectors $\textbf{w}_{kq}\;\forall k,\forall q$ are computed once in each coherence time interval, at the time instant $\lambda$, we can drop the time index for simplicity.

\par The normalization parameter $\mu_q$ is necessary to guarantee that each AP satisfies its maximum transmit power constraint $p_d$, and is computed as
    $1/{\mu_q} = \sum_{i=1}^K\mathbb{E}\{\lVert\textbf{w}_{iq}\rVert^2\}$.

\par The received signal at the $k$-th UE can be written as
\begin{equation}
    \begin{split}
        y_k[n]=&\underbrace{\sum_{q=1}^Q \sqrt{p_d\mu_q}\,\, \textbf{h}_{kq}^H[n] \,\, \textbf{w}_{kq}s_{k}[n]}_\text{Desired signals} + \\ &\underbrace{\sum_{i=1,i\neq k}^K\sum_{q=1}^Q \sqrt{p_d\mu_q}\,\, \textbf{h}_{kq}^H[n]\,\, \textbf{w}_{iq}s_{i}[n]}_\text{Inter-User Interference} + \underbrace{z_k[n]}_\text{Noise}.
    \end{split}
\end{equation}

\par In this work, we adopt the Use-and-then-Forget (UatF) bound technique to obtain semi-analytical expressions for the achievable downlink rates \cite{emil2017}. The UatF bound provides us with a rigorous lower bound regardless of the amount of channel hardening. Nevertheless, the more channel hardening we have on the system, the tighter this bound is. In other words, the UatF bound is a pessimistic estimate of the achievable downlink rate when channel hardening does not occur \cite{interdonato2019}.

\par Thus, the ergodic downlink capacity of the $k$-th UE is lower bounded using the UatF bound as
%
\begin{align}
    \label{eq:se-uatf}
    \texttt{SE}_k &= \dfrac{1}{\tau_c}\sum_{n=\lambda+1}^{\tau_c} \log_2(1+\gamma_k[n]) \text{ bits/s/Hz}, \\
    \label{eq:sinr-uatf}
    \gamma_k[n]& = \dfrac{p_d\bigg|\sum\limits_{q=1}^Q\texttt{DS}_{kq}[n]\bigg|^2}{p\sum\limits_{i=1}^K\texttt{INT}_i[n]-p_d\bigg|\sum\limits_{q=1}^Q\texttt{DS}_{kq}[n]\bigg|^2+\sigma^2}, \\
    \label{eq:DS-uatf}
    \texttt{DS}_{kq}[n] &= \mathbb{E}\{\sqrt{\mu_q}\textbf{h}_{kq}^H[n]\textbf{w}_{kq}\}    \\
    \label{eq:INT-uatf}   
    \texttt{INT}_i[n]&=\mathbb{E}\biggl\{\bigg|\sum_{q=1}^Q\sqrt{\mu_q}\textbf{h}_{kq}^H[n]\textbf{w}_{iq}\bigg|^2\biggl\},
\end{align}
where \eqref{eq:sinr-uatf} is the effective SINR at the $k$-th UE and at time instant $n$, \eqref{eq:DS-uatf} corresponds to the desired signal terms while and \eqref{eq:INT-uatf} corresponds to the inter-user interference terms.

\par Considering Maximum Ratio Transmission (MRT) precoding, we have
    $\textbf{w}_{kq}=\hat{\textbf{h}}_{kq}[\lambda],\;\forall k,\forall q$.    
Note that we consider the impact of imperfect CSI in the precoding vectors.


\section{Numerical Results}
\label{numericalResults}

\par In this section, we present Monte Carlo simulation results that illustrate the trade-off between macro-diversity and beamforming gains on DmMIMO. We fix the total number of antenna elements $M$ and the total downlink power $P$, and then we evaluate the trade-off between $Q$ and $S$, respectively the number of APs and antenna elements on each AP. The considered simulation parameters are listed in Table \ref{tableParameters}. 



\par Given the numbers of APs $Q$, antennas per AP $S$, and UEs $K$, we generate 50 network realizations, which consist of different sets of uniformly distributed positions for the UEs. Then, for each network realization, we generate 1000 channel realizations. The SE expressions are averaged over all the network and channel realizations. Then, the mean per-user achievable SE is obtained by averaging over the achievable SEs of all the $K$ UEs. The wrap-around technique is utilized to avoid the network border effect \cite{fastenbauer2019}.

\begin{table}[t!]
    \caption{Simulation parameters \cite{emil2019,ozdogan2019,tr25.996}.}
    \label{tableParameters}
    \centering
    \begin{tabular}{l c c}
        \toprule
        \textbf{Parameter} & \textbf{Symbol} & \textbf{Value}\\
        \midrule
        Number of UEs & $K$ &  20\\
        Number of APs & $Q$ &  \{1,4,16,64\}\\        
        Number of antennas on each AP & $S$ & \{64,16,4,1\}\\
        Total number of antenna elements & $M$ & \{64,128\}\\
        Length of the side of the square area & $l$ & \{125,250,500,1000\} m\\
        Signal bandwidth & $B$ & 20 MHz\\
        Coherence bandwidth & $B_c$ & 100 kHz\\
        Coherence time & $T_c$ & 2 ms\\
        Carrier frequency & $f_c$ & 2 GHz\\
        Sample time interval & $T_s$ & $10\;\mu\text{s}$\\
        Uplink transmit power & $p_k$ & 23 dBm\\
        Downlink total transmit power & $P$ & 49.03 dBm\\
        Noise figure at the UEs & $N_F$ & 9 dB\\
        Noise power & $\sigma^2$ & $-92$ dBm\\
        Coherence block & $\tau_c$ & 200 samples\\
        Length of the pilot sequence & $\tau_p$ & 10 samples\\
        Height of the APs & $h_{\text{AP}}$ & 12.5 m\\
        Height of the UEs & $h_{\text{UE}}$ & 1.5 m\\
        \bottomrule       
    \end{tabular}
\end{table}


\par
\begin{figure}[t]
    \centering
    \begin{subfigure}[t]{0.5\textwidth}
        \centering
        \includegraphics[scale=0.48]{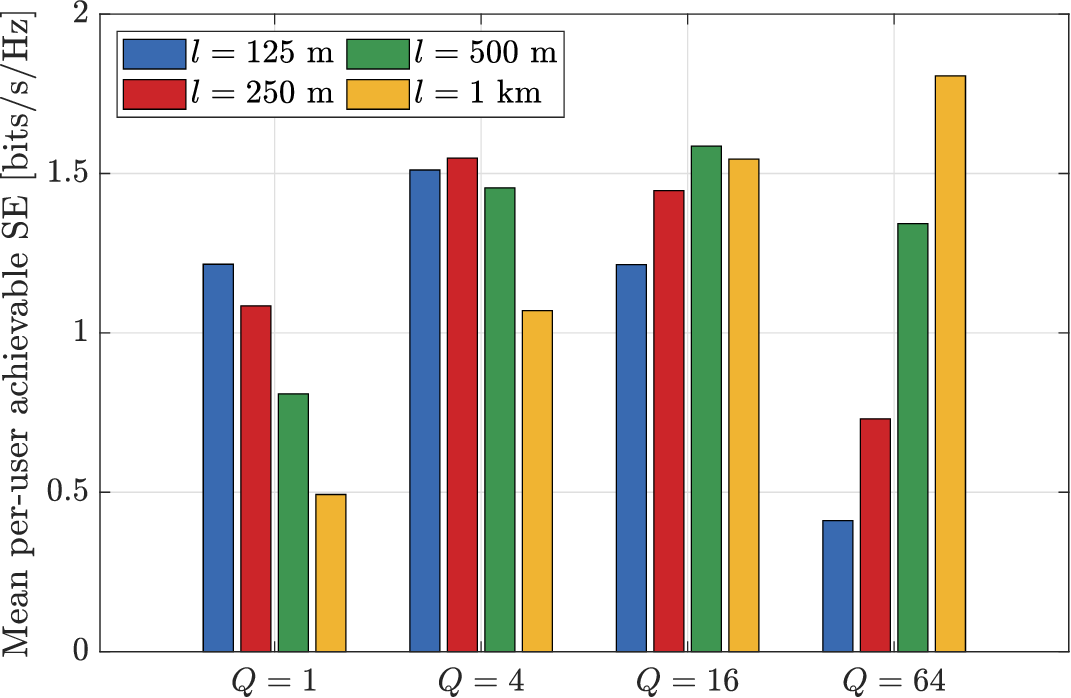}
        \caption{}
        \label{resultsNumberAPsM64}
    \end{subfigure}
    
    \begin{subfigure}[t]{0.5\textwidth}
        \centering
        \includegraphics[scale=0.48]{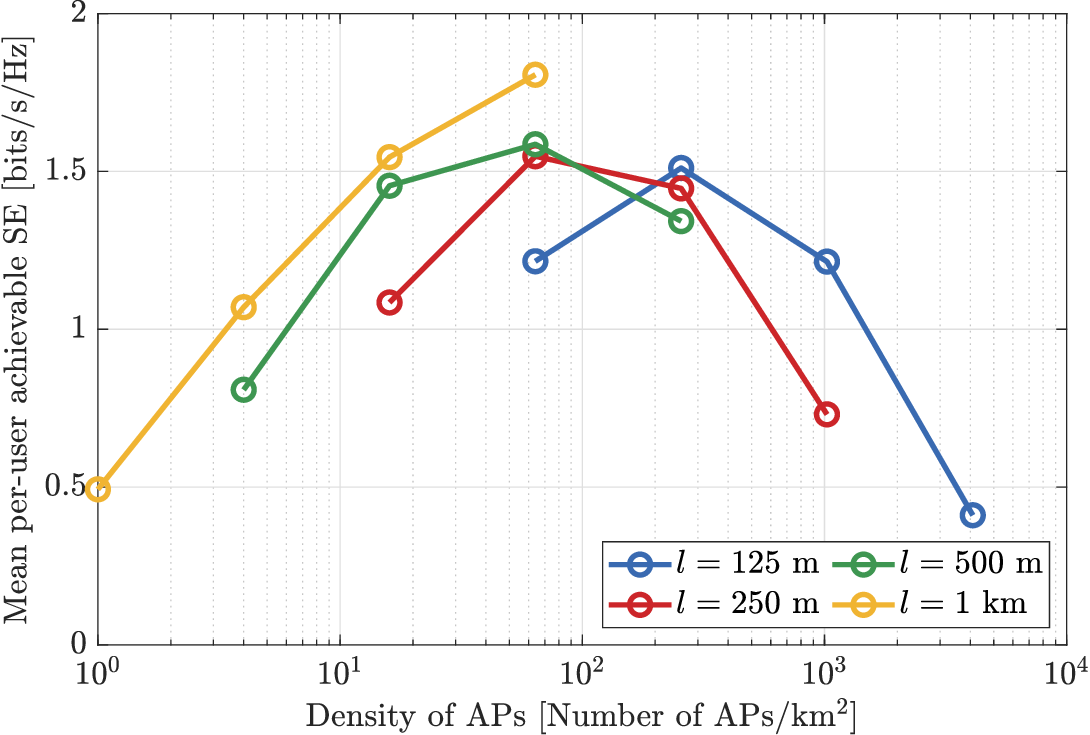}
        \caption{}
        \label{resultsDensityAPsM64}    
    \end{subfigure}

    \caption{Mean per-user achievable SE versus the number of APs (a) and density of APs (b) for $M=64$.}
    \label{resultsM64}
\end{figure}

\begin{figure}[t]
    \centering
    \begin{subfigure}[t]{0.5\textwidth}
        \centering
        \includegraphics[scale=0.48]{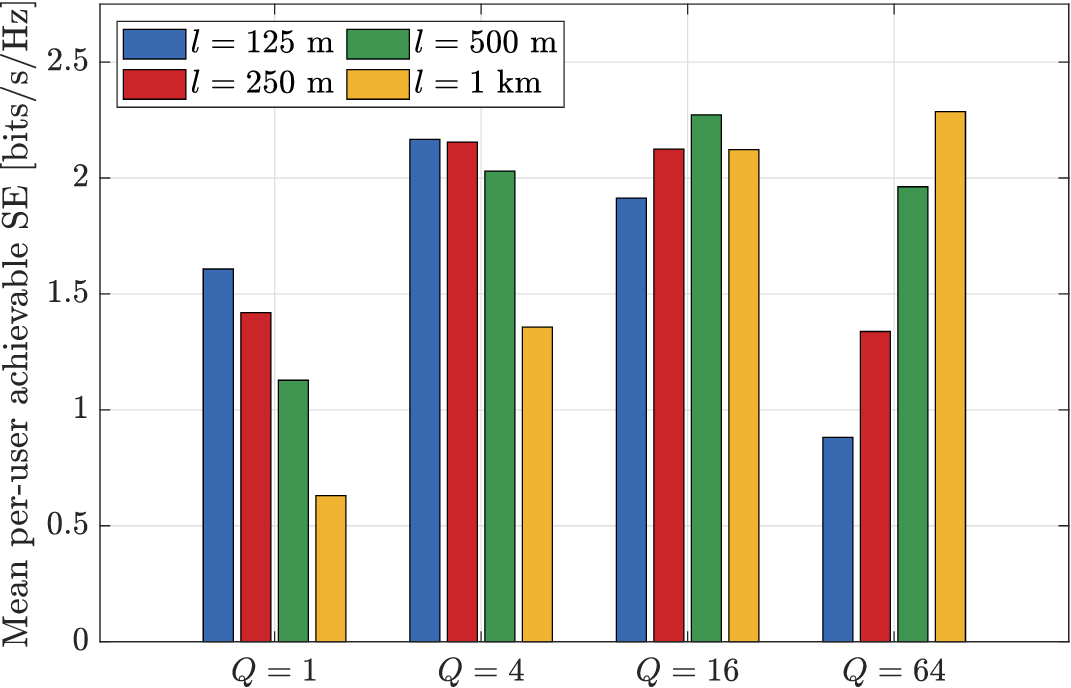}
        \caption{}
        \label{resultsNumberAPsM128}
    \end{subfigure}
    
    \begin{subfigure}[t]{0.5\textwidth}
        \centering
        \includegraphics[scale=0.48]{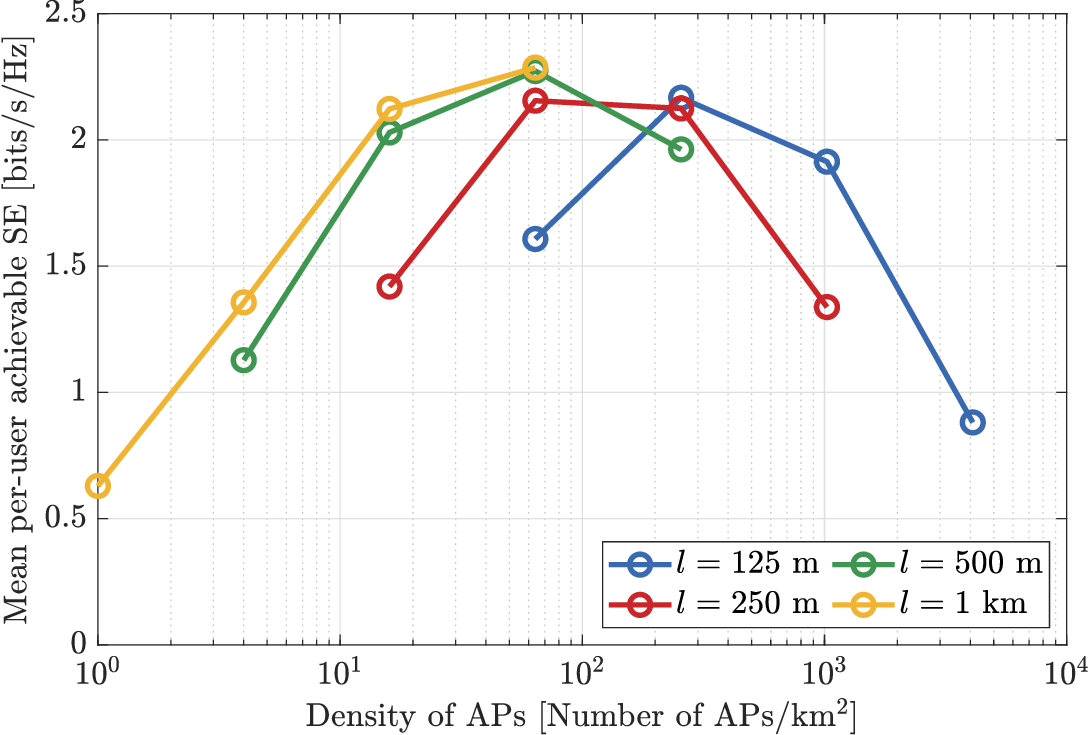}
        \caption{}
        \label{resultsDensityAPsM128}    
    \end{subfigure}

    \caption{Mean per-user achievable SE versus number of APs (a) and density of APs (b) for $M=128$.}
    \label{resultsM68}
\end{figure}

 Herein, all the numerical results were obtained adopting the LMMSE channel estimator. We present the mean per-user achievable SE for $M=64$ in Fig. \ref{resultsM64} and for $M=128$ in Fig. \ref{resultsM68}, for different number of APs $Q$, and also for different coverage areas $l^2$. The results are a function of the number of APs in Fig. \ref{resultsNumberAPsM64} and Fig. \ref{resultsNumberAPsM128} and the density of APs $Q/l^2$ in Fig. \ref{resultsDensityAPsM64} and Fig.\ref{resultsDensityAPsM128}. Note the optimal number of APs is a function of the dimensions of the coverage area.

\par For both the cases of $M=64$ and $M=128$, we observe that if the coverage area is relatively small ($l=125$ m or $l=250$ m), the best performance is achieved for the case of $Q=4$ APs. This shows that transitioning from a CmMIMO setup with $Q=1$ to a DmMIMO setup with $Q>1$ provides some macro-diversity gain that significantly improves the system's performance. Nevertheless, since the distances between the UEs and APs tend to be small, there are very high probabilities for the existence of LoS components between the UEs and APs; thus, the path loss is not severe. In this situation, the performance is also greatly improved by having some level of beamforming gain, i.e., having few APs equipped with multiple antenna elements is more advantageous than having many APs equipped with few or single antennas. In other words, for small coverage areas, there is a balance between the macro-diversity gain obtained from the spatial distribution of APs and the beamforming gain by having multiple antennas on each AP that guarantees the best performance. On the other hand, as the coverage area's size grows, the macro-diversity gains obtained become increasingly more important since the distances between UEs and APs are also increased. As a consequence, the probability of the existence of an LoS component between the UEs and APs becomes very low, and the path loss severely affects the performance. Distributing the antenna elements more sparsely on the coverage area increases the probability that a given UE is closer to at least one AP, thus increasing the probability of LoS. This is illustrated by the fact that the best performance for the cases of $l=500$ m and $l=1$ km is achieved with $Q=16$ and $Q=64$, respectively.

\par For the case of $Q=1$, increasing $l$ decreases the mean achievable per-user SEs, as expected, since the distances between the UEs and the single AP are also increased and the path losses are higher. Interestingly, for $Q=64$, increasing the size of the coverage area increases the performance. This happens because, as we increase the distances between the UEs, we reduce the levels of inter-user interference seen by each UE. The case of $Q=64$ and $l=1$ km could be interpreted as a small cell deployment \cite{ngo2017} since each UE tends to be close to only one AP and far away from the others.


\par Overall, given a total number of antenna elements $M$ and a total downlink power $P$, setups with few or a single AP equipped with multiple antenna elements present the best performance in small coverage areas. On the other hand, in bigger coverage areas, the UEs tend to be more sparsely distributed, and consequently, the path losses are more severe. Thus, macro-diversity gains become more beneficial than the beamforming gains, and having more APs equipped with few or single antenna elements is more advantageous.

\par The curves in Fig. \ref{resultsDensityAPsM64} and Fig. \ref{resultsDensityAPsM128} show that, for different values of $l$, the mean per-user achievable SE as a function of the density of APs $Q/l^2$ is a concave function. For both the cases of $M=64$ and $M=128$, the optimal density of $Q/l^2$ is approximately 100 APs/$\text{km}^2$.

\section{Conclusions}
\label{Conclusions}

\par In this work, we studied the trade-off between beamforming and macro-diversity gains on DmMIMO. Fixing the total number of antenna elements and the total downlink transmit power, we found that the ``sweet spot" on the number of APs and antenna elements per AP depends on the physical dimensions of the coverage area. If the UEs tend to be closer to the APs, beamforming gains provide more performance improvements than macro-diversity gains, i.e., it is better to have fewer APs equipped with multiple antennas. Conversely, if the distances between UEs and APs become longer, having more APs equipped with few or single antennas becomes advantageous since it increases the probability that a given UE is very close to at least one AP, thus having a high probability of LoS. 

\par Our numerical results show that DmMIMO networks that employ a moderate density of APs (on the order of 100 APs/$\text{km}^2$), with each AP equipped with multiple antenna elements, are the best option. Considering also that each AP requires a fronthaul connection to the CPU, fully distributed setups with a very high density of APs, i.e., setups that employ a massive number of single-antenna APs, may not be an economically viable option. Besides presenting worst performance in terms of mean per-user average SE, they also present very high deployment and maintenance costs. The results presented in this work can help network operators plan and deploy DmMIMO networks that simultaneously present reasonable deployment and maintenance costs and guarantee the best performance for the users.

\section*{Acknowledgment}

\par This research has been financially supported by Nokia Bell Labs, Academy of Finland, 6Genesis Flagship (grant no. 318937), European Union’s Horizon 2020 research and innovation programme (EU-H2020), Hexa-X-I (grant no. 101015956) and Hexa-X-II (grant no. 101095759) projects and CNPq (Brazil).

\bibliographystyle{./bibliography/IEEEtran}
\bibliography{./bibliography/references}

\end{document}